# The wireless router based on the linux system

**Abstract** – With the expansion of computer networks, the mobile terminal with wireless access capability experience a sharp increase in the number of wireless routers, especially low-cost wireless routers are becoming very important network equipment. This paper designs a wireless router based on the ARM platform, the Linux system. First, there is a research and analysis on the working principle and implementation of Network Address Translation (NAT) technology. Then I study the IPTABLES components under the Linux system and use it when processing data packets which go into the chain and the table and finally using laptop with Ethernet card and the WIFI card to build the Linux operating system. Related routing forwarding rules are defined between the two cards and use IPTABLES to achieve a laptop as a wireless WIFI hotspot providing routing and network connections to other computer services. It proves this paper's design, and production feasibility. The paper also discusses the design and production method of the wireless router with ARM board feasibility

**Keywords** –Network Address Translation; ARM platform; IPTABLES components

## 1. Introduction

In the past several years, the embedded technique was produced in the industry, computer networks, and consumption electronics. It got such a fierce development in our daily life. The embedded system was called as "taking the application as the core part, an specified computer system with a highly restrictive on the software, its credibility, cost, physical volume and power. Because the embedded system was extensively applied, as the soul of the embedded system, its software occupies a more and more important position in the whole software industry. The Internet which takes the network equipment as the basic constituent has become mature gradually in the recent years. Additionally the router is the most familiar network equipment, it will connect each subpart of the networks and finally become the whole Internet which may feel like a single unit network. A router is the hardware foundation of the construction of network which means that its function directly influence the function of the network. Currently, Internet has linked only 5% computer devices and a large quantity of embedded equipment was used to promote the service ability and application value of the network junction. However, based on the open source code, transplanting quickly, lowly requirement to the hardware, it can be enlarged as well as the strong support to the network, the embedded Linux operating system has got a very important position in the realm of embedded system. It has largely transplanted to the embedded equipment especially the network devices.

Of course the router is the most important node device among the Internet. The router decides the data forwarding by the router passes. Forwarded strategy can be called as the router choosing which is the original source of the word router. As the vital part of the connection of different network passes, the router system constructs the core part based on the international TCP/IP network. Its processing velocity has always been the bottleneck of the computer network and its credibility also influences the whole quality of the network connection. Therefore, in the technical village net, region net as well as in the whole Internet, router technique has always been the core part and its development and direction has become a microcosm of the whole Internet research. So in the current situation, discussion about the router strategy has become more and more meaningful[1].

Wireless router is like an expanded production which combines the wireless access point and the traditional router. It not only has the function of wireless access point as DHCP, supporting VPN, fire wall, WEP encryption, but also include the function of network address translation which refers to NAT and support the network share of local area network. It can realize the Internet connection in our houses as well as the ADSL and the small area network connection. The wireless router can be directly connected with all the Ethernet of ADSL MODERN and CABLE MODERN. There is a simple visual connection software inside can store the ID and password to connect the Internet. It can also automatically connect to the Internet through ADSL and CM instead of doing it by hands. In addition, wireless router has also contained the capability of a more complete security system.

## 2. ARM, LINUX and My router

First, let's talk about the ARM. ARM refers to the Advanced RISC Machine, which is the first RISC microcomputer that society faced. It was called Acorn RISC Machine, which is created by the Acorn Company. ARM is 32-bit design and it saves 35% of coding area compared to the traditional 32 bit Machine[2].

There are three basic merits of ARM which is low power consumption, 16/32-bit double instruction and a large amount of cooperation. We can get some additional detail by the following description:

(1) The small physical volume, low power, low cost and high capability.

(2) Support Thumb (16 bit) /ARM (32 bit) double instruction sets.

(3) Use the registers largely and have a more impressive implementing velocity.

(4) A lot of operation is completed in the registers.

(5) The way of address finding is quite flexible.

(6) The fixed number of instruction length.

The next is the Linux system. The Linux system belongs to the Unix system. It was created in the 1991 by a Finland computer genuine called Linux Torvalds. Based on the Internet, this system has been maintained by all the computer fans from all over the world and it has become the most used operating system among the UNIX subpart systems. In the meanwhile, the using population has been increased dramatically. From the birth of the Linux, based on its stability, safety, high performance and high expansion, it gets a high compliment of the users and become one of the most popular operating system gradually.

Linux acts like a free operating system in which customers can be free to use the original code and also can change the code. Because Linux has a huge number of similarities with the function of Unix, it will not need a high cost of Unix. What we have to know is that the price for Unix is even higher than Windows, and it's thousands of times higher than Linux.

The several parts construct the Linux system:

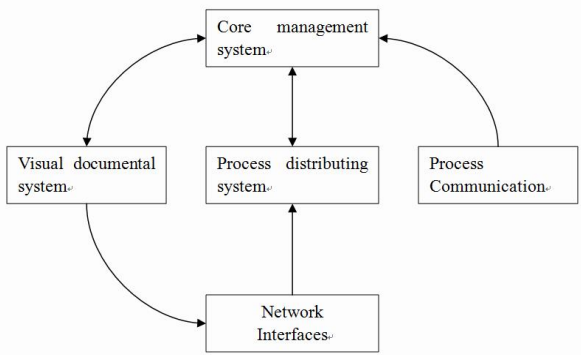

**Figure 1.** Core management system

From Table 1, what we can see is that from the core management system, we can find some function from the SCI which provides some implementing from the user's space to the core. The process distributing system is used to implement the main part of process. The visual documental system is a very impressive part of Linux core because it contains an abstract object to the system document. The process communication supports the data and address communication of the process. What the interface does is that it gives all the processes the ports to passing the messages.

Another important issue relating to Linux is the document structure. The document structure is the organization method what the document deposits in the memory devices. It mainly focuses on the organization of directory and catalog. The catalogue gives us a convenient and simple approach to manage the document. The users can switch from a directory to another directory. They can also set the limitation to use the document and the degree of sharing it.

Sometimes what we can find is that the mutual connection of document makes sharing the data to be easier and easier. We can simply find that multiple users can use the same document simultaneously. Linux is a multi-user's operating system, the main processes store in the specified directory that starts with the main directory. We can also see it from Table 2. Core, Shell and multiple documents comprise the basic operating system structure which makes the user able to implement process, manage document and use the whole system[3].

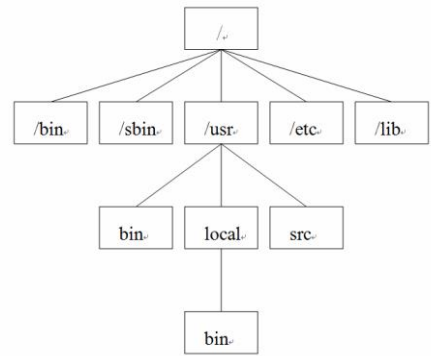

**Figure 2.** Elements of inputs, outputs, switches and router processor

Regarding to router, it was consisted of four elements, which are inputs, outputs, switches and router processor.

## 3. The basic protocol and skill about router

First is the TCP/IP protocol. It refers to Transmission control protocol/Internet Protocol. It is the most basic protocol and it is the fundamental part of the Internet network. What TCP/IP decides is that how electronic devices can connect to Internet and the standard rules for them to transfer data. It has four layers, each of the layer will call the next layer to complete its requirement. A traditional way to define is that TCP take charge of finding the transmission problems and solve all the problems until all things works well. IP means that each computer can only have one address[3].

**Table 1** TCP/IP protocol

| TCP/IP and OSI | |
|---|---|
| TCP/IP | OSI |
| application | application |
| | presentation |
| | sesstion |
| transport | transport |
| internet | internet |
| link | Data link |
| | physical |

From Table 3, what we can directly see is that from the respect of different layers of the model, TCP/IP consists of four layers: application, transport, Internet and link layers. Of course that the TCP/IP does not behave like the traditional OSI 7 layers..

## 4. NAT and IPTABLES

It is well known that the basic function of the router is to implement NAT, which refers to network address translation. NAT is the standard method that reflects an address to another address. It is based on the IETF standard of the opening of RFC1631. It allows a private IP to act like a public IP on the Internet. NAT can change all the addresses of local network to a public IP and it can also use the firewall to hide some private IP, which means that the users in the public network can not get the resources from the local network. So basically, NAT acts as a function of use a local network to visit a public network. This kind of IP translation can alleviate the situation of insufficient IP address area.

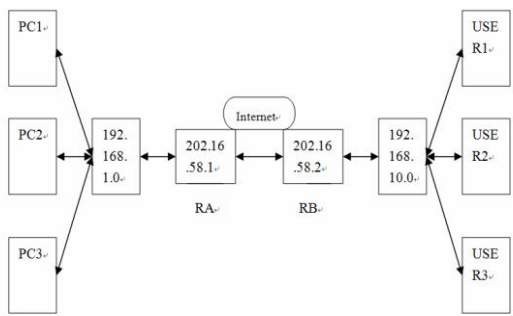

**Figure 3.** NAT working diagram

From Table 4, it is clear that the local private network can never visit the public network and we have to use NAT. So the table I show is like a network communication of two companies.

The IP of 192.168.1.0 wants to visit the user1 of 192.168.10.0.

(1) PC1 has to ask for requirement from RA1 to inform the private IP and MAC address. And also tell that it wants to visit 192.168.10.0.

(2) When RA1 get the requirement, it translates the private IP of PC1 to the public IP, which refers to 202.16.58.1. And it also randomly gives a port number to PC1, which can recognize the specific PC.

(3) The Internet network receives the claim of the translation and does the router selection which will be received by RB.

(4) The router RB gets the message. Based on the objective information, it translates the data to User1 from the network of 192.168.10.0.

(5) So based on the ICMP protocol, User1 need to make a respond. It needs to process the data and send to router after the data encapsulation.

(6) The router should translate the private IP to the public IP which is 202.16.58.2. And from this public IP, it can be sent to RA.

(7) When RA gets the package, it will check the PC and the port from the cache. And then it will send the data to PC1.

So here, in the Linux system, the ordinary way to construct a platform to make the system has the function of NAT is to use the IPTABLES. There are a lot of tables and chains in the IPTABLES and how to use this tables and chains to write the code to Linux is the approach to construct the router platform.

Netfilter/Iptables constitute the filtering firewall of Linux and it is free for sure. It can also replace the expensive business firewall solution and complete all the function like packaging, encapsulation, revising direction and network address translation. The IPTABLES module is also like a tool which can be called the userspace. It makes the rule of insertion, replacing, the deletion to be much easier. Unless you are using the Red Hat Linux 7.1 or higher edition, you have to download the tool to use it. Of course there three chains we have to use which is PREROUTING, POSTROUTING and OUTPUT[4].

So next, we will use the IPTABLES to implement NAT:

```
[root@localhost root]# iptables -t nat -A POSTROUTING -s 192.168.0.2 -o eth1 -j MASQUERADE
```

**Figure 3.** The IPTABLES to implement NAT

So from here, we can see that –o eth1 is to match the entire ip package to eth1. –j MASQUERADE means that we will transform the original address of the matched package to the IP address which corresponds to eth1.

Based on this order, we will complete the NAT translation from the local network to public network and give the visiting data package from the public network a legal original address. In the meanwhile, when we using the chain of POSRTOUTING to do the NAT, we will get a form, which will keep track of the relation between the WAN, ports to local IP address.

Actually, the wireless router is the wireless AP plus the router. So if we want to solve the problem, we should take care of the AP part.

AP refers to access point. It is the most common devices to construct a small wireless LAN. AP acts like a bridge from wire to wireless that connects the entire wireless network together and connect them all to the Ethernet.

So in this project, I use the USB net card from MERCURY to act as an AP. The edition of this net card is MW150UM which the 11N network rule. The velocity of wireless transforming can be 150M which can make the transforming to be more efficient and reduce the latency. Then I download a driver of this net card and make it to be checkable to Linux.

## 5. Summary and expectation.

So in this project, based on ARM, I mainly use the Linux system to get a deeper research and design the wireless router which implements NAT. I also talk about the IP protocols. And last I discuss the NAT and IPTABLES and use the IPTABLES to implement the function.

So in the next few years, the development intention of the router devices will still be of the high I/O velocity, higher switching ability. It will still focus on the service development, security and service condition. In 5 to 10 years, we will be facing a revolution in the Internet. Currently, we are facing a bottleneck that the speed rate can not increase as fast as we expect. Therefore, there must be a huge revolution to the industry about the electrical devices. So that if the router wants to survive in the revolution, it must be adaptable to reality or it may be replaced by totally new devices.